\documentclass{pasj00}
\begin{document}
\SetRunningHead{Matsuoka et al.}{Near-IR Spectroscopy of {\it IRAS} Galaxies}
%\Received{}%{yyyy/mm/dd}
%\Accepted{}%{yyyy/mm/dd}
%\Published{}%{yyyy/mm/dd}

\title{OAO/ISLE Near-IR Spectroscopy of {\it IRAS} Galaxies}

%%% begin:list of authors
% Do NOT capitalize all letters in "textsc".
\author{Y. \textsc{Matsuoka}, F.-T. \textsc{Yuan}, Y. \textsc{Takeuchi}} %
\affil{Graduate School of Science, Nagoya University, Furo-cho, Chikusa-ku, Nagoya 464-8602, Japan.}
\email{matsuoka@a.phys.nagoya-u.ac.jp}
\and
\author{K. \textsc{Yanagisawa}}
\affil{Okayama Astrophysical Observatory, National Astronomical Observatory of Japan, Kamogata, 
  Asaguchi, Okayama 719-0232, Japan.}
%%% end:list of authors

%%% Please use the following style in case that sorting by 
%%% affiliation is impossible. 
%
% \author{%
%   D-Firstname \textsc{D-Familyname}\altaffilmark{1}
%   E-Firstname \textsc{E-Familyname}\altaffilmark{1,2}
%   and
%   F-Firstname \textsc{F-Familyname}\altaffilmark{2}}
% \altaffiltext{1}{Address of Institute}
% \email{ddddd@xxx.xxx.xx.xx}
% \email{eeeee@xxx.xxx.xx.xx}
% \altaffiltext{2}{Address of Institute}

%% `\KeyWords{}' always has to be placed before `\maketitle'.
 %Do NOT move this preamble from here!

\maketitle

\begin{abstract}
We present the results of the near-infrared (IR) spectroscopy of nine {\it IRAS} galaxies (NGC 1266, NGC 1320, NGC 2633, NGC 2903, NGC 3034,
Mrk 33, NGC 7331, NGC 7625, NGC 7714) with the ISLE imager and spectrograph mounted on the Okayama Astrophysical Observatory 
1.88 m telescope.
[Fe II] 1.257 $\mu$m and Pa$\beta$ emission lines were observed for the whole sample while H$_2$ 2.121 $\mu$m and Br$\gamma$ lines
were additionally obtained for two sources, whose flux ratios are used as a diagnostic tool of dominant energy sources of the galaxies.
We find that the nucleus of NGC 1266 is most likely a low ionization nuclear emission-line region (LINER), while NGC 2633 
and NGC 2903 possibly harbor active galactic nuclei (AGNs).
No AGN or LINER signal is found for other objects.
In addition, we find the spectral features which is indicative of some unusual phenomena occurring in the galaxies, such as
the large [Fe II] line widths compared to the local escape velocity in NGC 1266.
The present work shows the potential ability of the ISLE to shed new light on the nature of infrared galaxies, either through 
a statistical survey of galaxies or an exploration of spectral features found in individual objects.
\end{abstract}

\section{Introduction}

One of the major achievements of the {\it Infrared Astronomical Satellite} ({\it IRAS}) is the discovery of a significant number of
galaxies emitting the bulk of their radiation at far-infrared (IR) wavelengths.
Luminous IR galaxies (LIRGs, with IR luminosity $L_{\rm IR} > 10^{11} L_{\odot}$) are as numerous as optical starburst and Seyfert 
galaxies with similar bolometric luminosity in the local universe,
while the more extreme class of ultra-luminous IR galaxies (ULIRGs; $L_{\rm IR} > 10^{12} L_{\odot}$) has similar space density
and bolometric luminosity to optically-selected quasars \citep{soifer87}.
Many studies have been devoted to the exploration of the origin of these IR galaxies, and it is now becoming
the general agreement that strong interactions and mergers of gas-rich galaxies are the trigger for the majority of them.
In the complete sample of {\it IRAS} 1 Jy ULIRGs, \citet{veilleux02} find that nearly 100 \% of IR galaxies show strong 
signs of tidal interaction.

On the other hand, there is still much debate about energy sources of IR galaxies.
Their IR luminosity can certainly come from dust reprocessing of radiations from starburst and/or active galactic nuclei (AGNs)
activity.
In this context, it is worth noting that ULIRGs and AGNs most probably have evolutionary connection.
It is suggested that major mergers of gas-rich galaxies first form a "cool" ULIRG dominated by dusty starburst, which is followed
by a "warm" ULIRG phase when an AGN turns on and starts to heat surrounding dust.
Then the AGN evolves into the dominant energy source and blows away the surrounding dust cocoon, leading to an optically-bright
quasar (e.g., \cite{lonsdale06}).
Therefore, observational investigation of energy sources in IR galaxies at various evolutionary stages is a key for understanding
the whole picture of galaxy evolution including AGN phase.
However, such investigation has been hampered by heavy dust obscuration inside IR galaxies.
Although numerous efforts have been devoted to the subject at various wavelengths 
(e.g., \cite{veilleux95,tran01,ptak03,farrah07,imanishi07}),
the results are still largely controversial: different studies give the inconsistent estimates of relative contributions from starburst
and AGNs for the same population.

In order to address the issue of energy sources in IR galaxies, we aim to investigate the presence of AGNs independently from the
previous studies, in a large number of IR galaxies including ULIRGs, LIRGs, and less luminous populations.
In this paper, we present the near-IR spectra of nine {\it IRAS} galaxies with relatively low IR luminosity, obtained 
by the ISLE imager and spectrograph mounted on the Okayama Astrophysical Observatory (OAO) 1.88 m telescope.
Near-IR light has a few advantages over those at other wavelengths in probing IR galaxies.
First, it suffers much less from dust obscuration than optical light, since the extinction at the former wavelengths are only 
about a tenth of that at the latter wavelengths ($A_K \sim 0.1 A_V$).
Second, near-IR observations cost much less than X-ray, mid- or far-IR observations, 
%which need to be carried out with space telescopes, 
hence a statistical number of objects can easily be investigated.
Third, there are many emission lines suitable for probing their radiation sources in near-IR spectral region 
(e.g., \cite{matsuoka07,matsuoka08}).
\citet{veilleux99} demonstrate the power of near-IR spectroscopy by observing 39 ULIRGs and finding that at least 50 \% 
of optically-classified Seyfert 2 galaxies present hidden broad-line-region (BLR) emissions in near-IR lines.
It means that nuclear regions of many optical Seyfert 2 galaxies, obscured completely in optical wave bands, are actually optically 
thin at near-IR wave bands.

We observed the four near-IR emission lines, [Fe II] 1.257 $\mu$m and Pa$\beta$ for the whole sample of nine galaxies and
H$_2$ 2.121 $\mu$m and Br$\gamma$ for two of them, whose flux ratios have been proposed as a diagnostic tool of dominant energy 
sources of emission-line galaxies \citep{larkin98, rodriguez-ardila05}.
Note that this kind of diagnostic diagrams works only with a statistical number of sample, since there are always outlying objects.
In this sense, we do not intend to present conclusive arguments with the small number of objects observed in this work, but rather 
aim to demonstrate the capability of the ISLE to investigate near-IR emission lines of IR galaxies for future projects.
Hereafter, [Fe II] 1.257 $\mu$m and H$_2$ 2.121 $\mu$m lines are abbreviated to [Fe II] and H$_2$ for simplicity.

\section{Observation and Data Reduction}

The target objects were selected from the Imperial {\it IRAS}-FSC Redshift Catalog (IIFSCz; \cite{wang09}) based on
the {\it IRAS} Faint Source Catalog (FSC).
The observability of the four emission lines, [Fe II], Pa $\beta$, H$_2$, and Br $\gamma$, in the near-IR
atmospheric window gives a severe restriction on redshifts, while the additional constraints come from positions 
and near-IR brightness ($J < 12$ mag).
We summarize the redshifts, the Two Micron All Sky Survey (2MASS; \cite{skrutskie06}) $J$-band magnitudes,
the 60-$\mu$m fluxes ($f_{\rm 60 {\mu}m}$) and the IR luminosity ($L_{\rm IR}$) based on the {\it IRAS}
measurements, and the corresponding {\it IRAS} names of the targets in Table \ref{tab:targets}.
They are not extremely luminous at IR wavelengths compared to LIRGs or ULIRGs.

\begin{table*}
  \caption{Targets Summary}\label{tab:targets}
  \begin{center}
    \begin{tabular}{lrrrrl}
      \hline
           &          & $J$   & $f_{\rm 60 {\mu}m}$ & log $L_{\rm IR}$ & \\
      Name & Redshift & (mag) & (Jy)                & ($L_{\solar}$)   & {\it IRAS} name \\
      \hline
      NGC 1266 & 0.007318 & 10.66 &    12.83 & 10.44 & F03135$-$0236 \\
      NGC 1320 & 0.008883 & 10.47 &     2.15 & 10.24 & F03222$-$0313 \\
      NGC 2633 & 0.007205 & 10.13 &    15.87 & 10.65 & F08425$+$7416 \\
      NGC 2903 & 0.001855 &  7.03 &    47.62 & 10.10 & F09293$+$2143 \\
      NGC 3034 & 0.000677 &  5.88 &  1217.26 & 10.41 & F09517$+$6954 \\
      Mrk 33   & 0.004769 & 11.61 &     4.68 &  9.71 & F10293$+$5439 \\
      NGC 7331 & 0.002722 &  7.16 &    32.08 & 10.29 & F22347$+$3409 \\
      NGC 7625 & 0.005447 &  9.98 &     9.33 & 10.25 & F23179$+$1657 \\
      NGC 7714 & 0.009333 & 10.84 &    10.36 & 10.66 & F23336$+$0152 \\
      \hline
    \end{tabular}
  \end{center}
\end{table*}

The observation was carried out with the ISLE, a near-IR imager and spectrograph mounted on the OAO 1.88 m telescope 
\citep{yanagisawa06,yanagisawa08}.
The instrument uses a HAWAII 1k $\times$ 1k array, which provides 4.3 $\times$ 4.3 arcmin$^2$ field-of-view with a pixel 
scale of 0.25 arcsec.
The useful wavelength intervals covered by the $J$, $H$, and $K$ filters are 1.11 -- 1.32 $\mu$m, 1.50 -- 1.79 $\mu$m,
and 2.02 -- 2.37 $\mu$m, respectively.
%The covered wavelength range is $J$ (1.11 -- 1.32 $\mu$m), $H$ (1.50 -- 1.79 $\mu$m), and $K$ (2.02 -- 2.37 $\mu$m) bands.
The slit length and orientation are fixed to 4 arcmin and east--west direction.

The observation journal is given in Table \ref{tab:obs}.
For each object we first carried out $J$-band observation, and then decided the priority of taking additional $H$- or $K$-band 
data based on the $J$-band spectrum.
As a result, $H$-band spectrum was taken for NGC 1266 and $K$-band spectra were taken for NGC 2633 and NGC 2903 (more details
are given in the following section).
The sky condition was sometimes non-photometric during the nights.
We used the 2.0-arcsec slit, which provides wavelength resolution of $R$ $\sim$ 1200, 1800, and 1000 at $J$, $H$, and $K$ band,
respectively.
The choice of the relatively wide slit was effective in reducing amplitudes of fringe patterns which often dominate 
measurement errors at red part of the spectra.
%The slit orientation was fixed to east--west.
The total exposure times were broken into individual exposures of 120 sec, and the objects were
offset along the spatial direction of the slit between adjacent exposures.
We observed A0V -- A0III stars for flux calibration immediately before or after the target exposures at similar airmass.

\begin{table*}
  \caption{Observation Journal}\label{tab:obs}
  \begin{center}
    \begin{tabular}{lcccc}
      \hline
             &      &      & Exp time &  Standard\\
      Target & Date & Band & (min)    &  star\\
      \hline
      NGC 1266 & 2010 Dec 09 & $J$ & 128 & HD 13936, HD 32996  \\
               & 2010 Dec 10 & $H$ & 64  & HD 13936, HD 32996  \\
      NGC 1320 & 2010 Dec 08 & $J$ & 80  & HD 31411  \\
      NGC 2633 & 2010 Dec 09 & $J$ & 68  & HD 55075  \\
               & 2010 Dec 09 & $K$ & 56  & HD 55075, HD 71906  \\
      NGC 2903 & 2010 Dec 10 & $J$ & 32  & HD 55075, HD 89239  \\
               & 2010 Dec 10 & $K$ & 64  & HD 89239  \\
      NGC 3034 & 2010 Dec 10 & $J$ & 16  & HD 55075, HD 89239  \\
      Mrk 33   & 2010 Dec 08 & $J$ & 80  & HD 92728  \\
      NGC 7331 & 2010 Dec 09 & $J$ & 32  & HD 211096, HD 13936\\
      NGC 7625 & 2010 Dec 10 & $J$ & 108 & HD 208108, HD 1439  \\
      NGC 7714 & 2010 Dec 09 & $J$ & 32  & HD 208108, HD 211096 \\
      \hline
    \end{tabular}
  \end{center}
\end{table*}

The data reduction was performed in a standard manner.
After dark subtraction and flat fielding, the sky emission was eliminated by subtracting an adjacent offset image.
Then the residual sky background was estimated from counts of nearby pixels in the spatial direction, and removed.
We extracted the target spectra within a 5.0-arcsec (20-pixel) aperture, whose size is close to the full width at zero
intensity of the point spread function, in all but the $J$-band images of NGC 7331 and NGC 1266 requiring
a larger 7.6-arcsec (30-pixel) aperture due to the poor seeing.
The apertures were centered on spatial peak positions of the spectra.
All the galaxies have the apparent sizes of $>$10 arcsec, hence the resultant spectra sample only their central regions.
We also note that the estimated sky backgrounds could contain the contributions from the outer galaxies.
Such a contamination would reduce the galaxy contributions in the extracted spectra, which might results in a little 
enhancement of the relative contribution of AGNs.
However, the signal-to-noise ratios of outer galaxy regions in our data are not high enough to quantify these contributions.
Wavelength calibration was achieved by referring to the Ar arc spectra obtained with the same instrument configuration 
as used for the target observation.
The mean RMS of the calibration are 0.9, 3.4, and 1.7 \AA\ in $J$, $H$, and $K$ band, respectively.
The atmospheric and instrumental transmission was estimated from the observed spectra of the A0 standard stars, for 
which we assumed an intrinsic black-body spectrum with the effective temperature of 9600 K \citep{pickles98}
after stellar H I recombination lines were manually removed.

\section{Results and Discussion}

\begin{figure*}
  \begin{center}
    \FigureFile(140mm,140mm){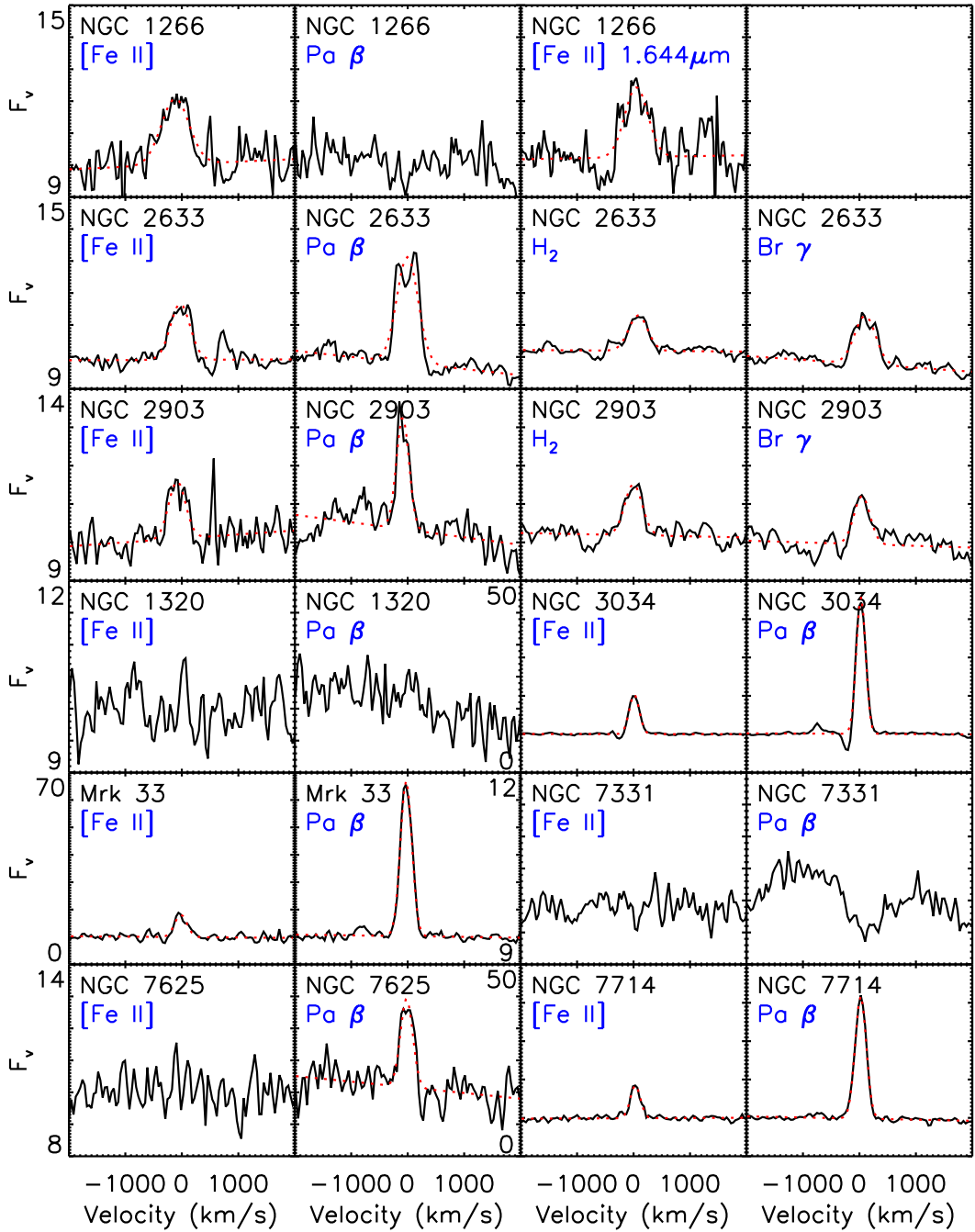}%{../observations/oao10dec/analysis/galspec.ps}
  \end{center}
  \caption{ISLE spectra of the {\it IRAS} galaxies around [Fe II] 1.257 $\mu$m, Pa $\beta$ (observed in $J$ band), 
    H$_2$ 2.1213 $\mu$m, and Br $\gamma$ (in $K$ band) emission lines, as well as [Fe II] 1.644 $\mu$m
    (in $H$ band) for NGC 1266.
    The wavelengths have been converted to the relative velocities to the line centers.
    The target name and the emission line are indicated at the top-left corner of each panel.
    The fitted Gaussian functions are also shown for the detected lines (red dotted lines).
    The fluxes are given per velocity in arbitrary scale, while those of a same object are shown in the same scale.}
  \label{spec}
\end{figure*}

We show the reduced spectra around the four emission lines, as well as [Fe II] 1.644 $\mu$m for NGC 1266, in Figure \ref{spec}.
To the detected lines, we fit Gaussian functions with underlying continua represented by tilted lines, i.e.,
\begin{equation}
  F (v) = a_0\ exp \left[-\frac{(v - a_1)^2}{2 a_2^2}\right] + (a_3 + a_4 v) ,
\end{equation}
where $v$ represents velocity shift relative to the line centers.
The free parameters $a_0$, $a_1$, $a_2$, $a_3$, and $a_4$ are determined simultaneously by fitting the function to
the observed spectra with the least-$\chi^2$ method within velocity range from $-$2000 to $+$2000 km s$^{-1}$.
We show the fitted functions in Figure \ref{spec}.
The equivalent widths (EWs) and the full widths at half maximum (FWHMs) of the emission lines derived from the 
best-fit parameters are summarized in Table \ref{tab:lines}.
The FWHMs were corrected for the instrumental resolution assuming $R$ = 1200, 1800, and 1000 in $J$, $H$, and $K$ band,
respectively.
The upper limit of 300 km s$^{-1}$ was given to the lines with the measured FWHMs less than this value.
The reduced $\chi^2$ values are close to one in all the cases.
Note that we use EWs and flux ratios rather than absolute flux values, since the latter is subject to the unknown amounts
of aperture losses.
% caused by misalignment of the slit on the target peak positions or by the variable seeing between the 
% target and standard star measurements.

\begin{table*}
  \caption{Measured Emission-Line Properties}\label{tab:lines}
  \begin{center}
    \begin{tabular}{lcccccccc}
      \hline
             & \multicolumn{2}{c}{[Fe II] 1.257 $\mu$m} & \multicolumn{2}{c}{Pa $\beta$} & \multicolumn{2}{c}{H$_2$ 2.121 $\mu$m} & \multicolumn{2}{c}{Br $\gamma$}\\
             & EW  & FWHM  & EW  & FWHM & EW & FWHM  & EW & FWHM \\
      Object & (\AA) & (km s$^{-1}$) & (\AA) & (km s$^{-1}$) & (\AA) & (km s$^{-1}$) & (\AA) & (km s$^{-1}$)\\
      \hline
      NGC 1266 & 5.3 $\pm$ 1.5 & 520 $\pm$ 60 & $<$ 0.6        & --           & --    & --    & --    & --   \\
      NGC 1320 & $<$ 0.4       & --           & $<$ 0.3        & --           & --    & --    & --    & --   \\
      NGC 2633 & 3.2 $\pm$ 0.5 & 320 $\pm$ 20 & 6.8  $\pm$ 0.5 & 360 $\pm$ 10 & 3.2 $\pm$ 0.5 & 260 $\pm$ 30 & 5.2 $\pm$ 0.5 & 310 $\pm$ 20 \\
      NGC 2903 & 2.3 $\pm$ 0.8 & 240 $\pm$ 40 & 3.4  $\pm$ 0.7 & $<$ 300      & 3.6 $\pm$ 0.7 & 200 $\pm$ 30 & 3.2 $\pm$ 0.8 & 140 $\pm$ 40 \\
      NGC 3034 & 9.6 $\pm$ 0.4 & $<$ 300      & 30.9 $\pm$ 1.3 & $<$ 300      & --    & --    & --    & --   \\
      Mrk 33   & 9.3 $\pm$ 1.7 & $<$ 300      & 59.7 $\pm$ 1.9 & $<$ 300      & --    & --    & --    & --   \\
      NGC 7331 & $<$ 0.2       & --           & $<$ 0.3        & --           & --    & --    & --    & --   \\
      NGC 7625 & $<$ 0.6       & --           & 3.2  $\pm$ 0.7 & $<$ 300 & --    & --    & --    & --   \\
      NGC 7714 & 8.1 $\pm$ 0.8 & $<$ 300      & 34.5 $\pm$ 0.9 & $<$ 300      & --    & --    & --    & --   \\
      \hline
             & \multicolumn{2}{c}{[Fe II] 1.644 $\mu$m} & \multicolumn{2}{c}{} & \multicolumn{2}{c}{} & \multicolumn{2}{c}{}\\
             & EW  & FWHM  &   &  &  &   &  &  \\
      Object & (\AA) & (km s$^{-1}$) &  & & & & & \\
      \hline
      NGC 1266 & 6.0 $\pm$ 1.5 & 450 $\pm$ 40 & & &   &   &   &   \\
      \hline
    \end{tabular}
  \end{center}
  Note --- 2-$\sigma$ upper limits are given for the fluxes of undetected emission lines.
\end{table*}

\begin{figure}
  \begin{center}
    \FigureFile(80mm,80mm){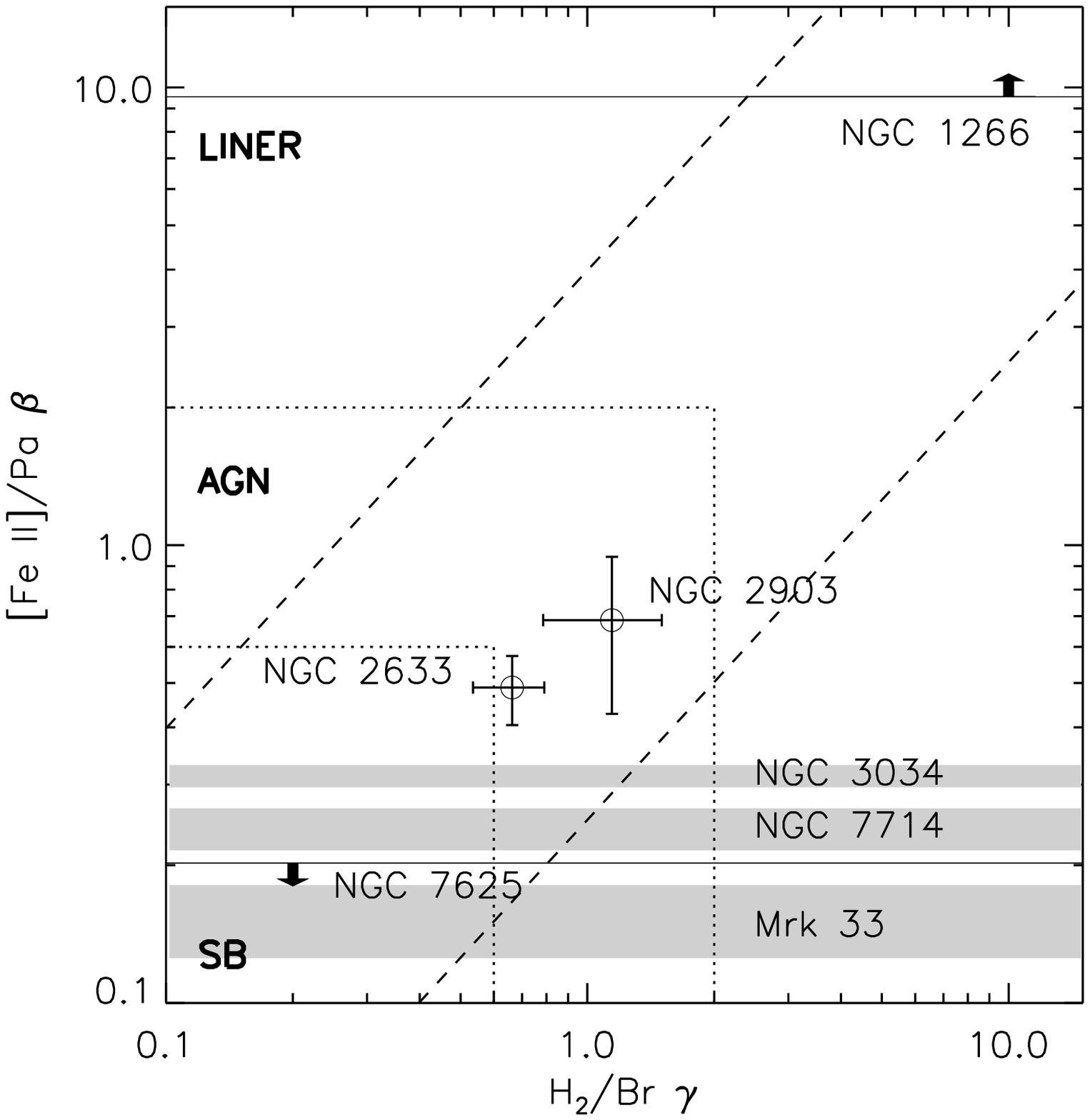}%{../observations/oao10dec/analysis/lineratios.ps}
  \end{center}
  \caption{Measured line flux ratios of [Fe II]/Pa $\beta$ and H$_2$/Br $\gamma$.
    The circles represent the objects with the four lines available, while the shaded area (1$\sigma$ confidence level) or
    horizontal lines with arrows (the up and down arrow shows lower and upper limit, respectively) 
    represent those without H$_2$ and Br $\gamma$ measurements.
%    We assume the arbitrary values of H$_2$/Br $\gamma$ = 0.2 or 10.0 for the latter objects in the plot.
    The object names are indicated near the corresponding symbols.
    The dotted lines show the starburst (SB)/AGN/LINER demarcation proposed by \citet{rodriguez-ardila05},
    while the dashed lines show the approximate envelope of distribution of the objects compiled by them.}
  \label{lineratios}
\end{figure}

The measured line flux ratios of [Fe II]/Pa $\beta$ and H$_2$/Br $\gamma$ are plotted in Figure \ref{lineratios}.
\citet{larkin98} suggest this diagram as a diagnostic tool for energy sources of the line emissions.
Later the classification scheme is updated by \citet{rodriguez-ardila04} and \citet{rodriguez-ardila05}, who find that
AGNs are characterized by the two ratios between 0.6 and 2, while the smaller or larger values indicate starburst/H II 
galaxies or low ionization nuclear emission-line regions (LINERs), respectively (dotted lines of Figure \ref{lineratios}).
However, note that this diagnostic diagram works only with a statistical number of sample, since there are the objects 
that do not meet the above criteria in the compilation of \citet{rodriguez-ardila05}.

The most noticeable object in Figure \ref{lineratios} is NGC 1266.
Its [Fe II] line flux is at least 10 times larger than Pa $\beta$, clearly indicating it is a LINER.
Therefore we obtained the additional $H$-band spectrum and found that another [Fe II] line,
[Fe II] 1.644 $\mu$m, is also strong in this object.
The FWHMs of the two [Fe II] lines, $\sim$500 km s$^{-1}$, are significantly larger than the local escape velocity 
($<$ 340 km s$^{-1}$; \cite{alatalo11}), which is indicative of an energetic phenomenon occurring in the galaxy.
Actually, the presence of an AGN or LINER in the galaxy is implied by the radio, optical, and X-ray observations
(\cite{alatalo11} and references therein).
A powerful molecular wind from the nucleus of NGC 1266 is found, which is suggested to be driven by an AGN since the
estimated star-formation rate is insufficient for the drive.
%The CO emission shows a broad wing component, more than 2.5 \% of whose velocities exceed the innermost local escape velocity.
Our near-IR exploration present a new evidence for the LINER nature of this galaxy.

NGC 2633 and NGC 2903 have the consistent [Fe II]/Pa $\beta$ and H$_2$/Br $\gamma$ ratios with AGNs.
$H$- and $K$-band spectra of NGC 2633 were obtained by \citet{vanzi98}, who find that H$_2$ and Br $\gamma$ are clearly 
detected while [Fe II] 1.644 $\mu$m is not.
The EWs of H$_2$ and Br $\gamma$ measured by them, 3.0 $\pm$ 0.5 \AA\ and 5.0 $\pm$ 0.5 \AA, are in excellent agreement with our 
results. 
While \citet{vanzi98} classify this galaxy as a starburst based on the optical [O I] 0.63 $\mu$m/H $\alpha$ and near-IR 
[Fe II] 1.644 $\mu$m/Br $\gamma$ line ratios, our new data indicate that NGC 2633 may be more like an AGN.
In reality NGC 2633 sits on the borderline of starbursts and AGNs in our diagram and of starbursts and "composite" objects
in the Vanzi et al. classification diagram, suggesting the contributions from both starburst and AGN activity.
NGC 2903 is a well-known starburst galaxy and has been studied at various wavelengths from radio to X-ray 
(e.g., \cite{popping10}).
While the majority of the radiation is thought to come from star-formation activity, \citet{perez-ramirez10} raise the 
possibility of a low-luminosity AGN present in this galaxy based on its X-ray property.
The near-IR line flux ratios presented here support this possibility.
%, but it is also possible that a faint signal of
%the low-luminosity AGN, if it exists, is obscured by the dominant contribution from starburst.
%In the latter case, NGC 2903 is an exceptional starburst galaxy located at AGN space in the [Fe II]/Pa $\beta$ -- H$_2$/Br $\gamma$ 
%diagram.

The [Fe II]/Pa $\beta$ ratios of NGC 3034, Mrk 33, NGC 7625, and NGC 7714 are found to be at most 0.3, while no observation is available
for the measurement of H$_2$ and Br $\gamma$.
In the sample compiled by \citet{rodriguez-ardila05}, all the sources with [Fe II]/Pa $\beta$ $\le 0.3$ have H$_2$/Br $\gamma$ $\le 0.3$.
It may imply that the above four objects are most likely categorized into starburst galaxies in Figure \ref{lineratios}, while the
additional observations of H$_2$ and Br $\gamma$ lines in these objects are needed for further discussion on this point.
%They are most likely categorized into starburst galaxies in Figure \ref{lineratios}, since all the objects with [Fe II]/Pa $\beta$ 
%$\le 0.3$ have H$_2$/Br $\gamma$ $\le 0.3$ in the compilation of \citet{rodriguez-ardila05}.
No clear evidence for the presence of AGNs has been obtained at other wavelengths, despite numerous observations dedicated to these 
well-known starburst galaxies.
Another worth-noting point is the P Cygni profile observed in [Fe II] and Pa $\beta$ lines of NGC 3034 (see Figure \ref{spec}).
The absorption features are detected at $\sim$250 km s$^{-1}$ blueshift relative to the line centers, pointing to the presence 
of outflowing material in front of the line emission region.
%In addition, another emission component is found at $\sim$770 km s$^{-1}$ blueward of Pa $\beta$ in NGC 3034.
These features are not found in the optical spectra of nuclear star clusters of this galaxy
(e.g., \cite{westmoquette07, konstantopoulos09}), which may be due to the heavy dust obscuration.
Observations with higher spatial resolution than the present one are required for locating this velocity component in the 
nuclear region of the galaxy.

\section{Conclusion}

In this paper we demonstrate the capability of the ISLE imager and spectrograph mounted on the OAO 1.88 m telescope to
investigate near-IR line emissions of {\it IRAS} galaxies.
The observation targets were selected from the IIFSCz based on the {\it IRAS} FSC.
All of them are the nearby galaxies at the redshifts $z < 0.01$. % with the $J$-band magnitudes brighter than 12 mag.
[Fe II] 1.257 $\mu$m and Pa $\beta$ emission lines were observed in a sample of nine galaxies, while H$_2$ 2.121 $\mu$m and Br $\gamma$ 
lines were additionally observed for two of them.
%of which seven have the detectable fluxes in either of the lines by the ISLE.
Based on the measured line flux ratios, NGC 1266 is found to have a LINER, while NGC 2633 and NGC 2903 possibly harbor AGNs.
On the other hand, no AGN signal is found for NGC 3034, Mrk 33, NGC 7625, and NGC 7714.
%These results are in general agreement with the indications obtained from the previous observations at other wavelengths.
In addition, we find the relatively large [Fe II] line widths of $\sim$500 km s$^{-1}$ in NGC 1266 and the P Cygni profile
%with an additional emission component at $\sim$770 km s$^{-1}$ blueward
of [Fe II] and Pa $\beta$ lines in NGC 3034.
The present work shows the potential ability of the ISLE near-IR spectroscopy to shed new light on the nature of IR
galaxies, either through a statistical survey of galaxies on diagnostic diagrams such as Figure \ref{lineratios}, 
or an exploration of spectral features found in individual galaxies.

%%%%%%%%%%%%%%%%%%%%%%%%%%%%%%%%%%%%%%%

\bigskip

This work was supported by Grants-in-Aid for Scientific Research (22684005) and the Global COE Program of Nagoya University 
"Quest for Fundamental Principles in the Universe" from JSPS and MEXT of Japan.

%%%
% See the manual for the detail.
%%%

\end{document}